\documentclass[usenatbib,usegraphicx,useAMS]{mn2e}

\usepackage{times}
\usepackage{amsfonts}
\usepackage{amssymb}

\newcommand{\fig}{Fig.\,}

\newcommand{\eq}{Eqn.\,}

\newcommand{\secref}{Section\,\ref}

\newcommand{\mpch}{\mbox{\,Mpc/$h$}}

\newcommand{\mbi}[1]{\mbox{\boldmath$#1$}}

\begin{document}
\title[High redshift galaxies]{The high redshift galaxy population in
  hierarchical galaxy formation models} 
\author[Kitzbichler \&
  White]{M.~G.~Kitzbichler\thanks{E-mail:~mgk@mpa-garching.mpg.de} and
  S.~D.~M.~White\\ Max-Planck Institut f\"ur Astrophysik,
  Karl-Schwarzschild-Stra\ss e~1, D-85748 Garching b. M\"unchen,
  Germany} \maketitle
\begin{abstract}
  We compare observations of the high redshift galaxy population to
  the predictions of the galaxy formation model of \citet{Croton2006}
  and \citet{Delucia2006b}. This model, implemented on the Millennium
  Simulation of the concordance $\Lambda$CDM cosmogony, introduces
  ``radio mode'' feedback from the central galaxies of groups and
  clusters in order to obtain quantitative agreement with the
  luminosity, colour, morphology and clustering properties of the
  present-day galaxy population. Here we construct deep light cone
  surveys in order to compare model predictions to the observed counts
  and redshift distributions of distant galaxies, as well as to their
  inferred luminosity and mass functions out to redshift 5. With the
  exception of the mass functions, all these properties are sensitive
  to modelling of dust obscuration. A simple but plausible treatment
  agrees moderately well with most of the data. The predicted
  abundance of relatively massive ($\sim M_*$) galaxies appears
  systematically high at high redshift, suggesting that such galaxies
  assemble earlier in this model than in the real Universe. An
  independent galaxy formation model implemented on the same
  simulation matches the observed mass functions slightly better, so
  the discrepancy probably reflects incomplete galaxy formation
  physics rather than problems with the underlying cosmogony.
\end{abstract}

\section{Introduction}
\label{sec:intro}
Recent work has used the very large Millennium Simulation to
follow the evolution of the galaxy population throughout a large
volume of the concordance $\Lambda$CDM cosmogony \citep{Springel2005b,
Croton2006,Bower2006,Delucia2006a,Delucia2006b}. By implementing
``semi-analytic'' treatments of baryonic processes on the stored
merger trees of all halos and subhalos, the formation and evolution of
about $10^7$ galaxies can be simulated in some detail. The inclusion
of ``radio mode'' feedback from the central galaxies in groups and
clusters allowed these authors to obtain good fits to the local galaxy
population and to cure several problems which had plagued
earlier galaxy formation modelling of this type. In particular, they
were able to produce galaxy luminosity functions with the observed
exponential cutoff, dominated at bright magnitudes by passively
evolving, predominantly elliptical galaxies. At the same time, this
new ingredient provided an energetically plausible explanation for the
failure of ``cooling flows'' to produce extremely massive galaxies in
cluster cores. Most of this work compared model predictions to the
systematic properties and clustering of the observed low redshift
galaxy population, or studied the predicted formation paths of massive
galaxies. Only \citet{Bower2006} compared their model in detail to
some of the currently available data at high redshift.  In the present
paper we compare these same data and others to the galaxy formation
model of \citet{Croton2006} as updated by \citet{Delucia2006b} and
made publicly available through the Millennium Simulation data
site.\footnote{http://www.mpa-garching.mpg.de/millennium; see
\citet{Lemson2006}}

Many recent observational studies have emphasised their detection of
substantial populations of massive galaxies out to at least redshift 2
and have seen this as conflicting with expectations from hierarchical
formation models in the $\Lambda$CDM cosmogony \citep[e.g.][]
{Cimatti2002c,Im2002,Pozzetti2003,Kashikawa2003,Chen2003,Somerville2004}.
This notion reflects in part the fact that early hierarchical models
assumed an Einstein-de-Sitter cosmogony in which recent evolution is
stronger than for $\Lambda$CDM \citep[e.g.][]{Fontana1999}, in part an
underassessment of the predictions of the contrasting toy model in
which massive galaxies assemble at high redshift and thereafter evolve
in luminosity alone \citep[see][]{Kitzbichler2006}. \citet{Bower2006}
find their model to be in good agreement with current observational
estimates of the abundance of massive galaxies at high redshift, while
our comparisons below suggest that the model of \citet{Croton2006} and
\citet{Delucia2006b} appears, if anything, to {\it overpredict} this
abundance. As shown by these authors and particularly by
\citet{Delucia2006a} both models predict ``anti-hierarchical''
behaviour, in that star formation completes earlier in more massive
galaxies. This behaviour clearly does not conflict with the underlying
hierarchical growth of structure in a $\Lambda$CDM cosmogony.

The current paper is organised as follows. In \secref{sec:model} we
briefly describe the Millennium Simulation and the fiducial galaxy
formation model we are adopting. Where we have made modifications,
most significantly in the dust treatment, these are described in
detail. We also give a detailed account of how we construct mock
catalogues of galaxies along the backward lightcone of a particular
simulated field of observation. Many of our methods resemble those
which \citet{Blaizot2005} implemented in their MOMAF facility in order
to enable mock observations of simulated galaxy catalogues of the same
type as (though smaller than) the Millennium Run catalogues we use
here. Our results are summarised in \secref{sec:results} where we
compare number counts as a function of apparent magnitude and redshift
with the currently available observational data. We also compare the
predicted evolution of the luminosity and stellar mass functions to
results derived from recent observational surveys, and we illustrate
how the population of galaxies is predicted to shift in the
colour-absolute magnitude plane. Finally in \secref{sec:discussion} we
interpret our findings and present our conclusions.

\section{The Model}
\label{sec:model}
\subsection{The Millennium dark matter simulation}
\label{darkmatter}
We make use of the Millennium Run, a very large simulation which
follows the hierarchical growth of dark matter structures from
redshift $z=127$ to the present. The simulation assumes the
concordance $\Lambda$CDM cosmology and follows the trajectories of
$2160^3\simeq 1.0078\times 10^{10}$ particles in a periodic box
500\mpch\ on a side. A full description is given by
\citet{Springel2005b}; here we summarise the main simulation
characteristics as follows:

The adopted cosmological parameter values are consistent with a
combined analysis of the 2dFGRS \citep{Colless2001} and the first-year
WMAP data \citep{Spergel2003,Seljak2005}. Specifically, the simulation
takes $\Omega_{\rm m}= \Omega_{\rm dm}+\Omega_{\rm b}=0.25$,
$\Omega_{\rm b}=0.045$, $h=0.73$, $\Omega_\Lambda=0.75$, $n=1$, and
$\sigma_8=0.9$ where all parameters are defined in the standard
way. The adopted particle number and simulation volume imply a
particle mass of $8.6\times 10^8\,h^{-1}{\rm M}_{\odot}$. This mass
resolution is sufficient to resolve the haloes hosting galaxies as
faint as $0.1\,L_\star$ with at least $\sim 100$ particles. The
initial conditions at $z\!=\!127$ were created by displacing particles
from a homogeneous, `glass-like' distribution using a Gaussian random
field with the $\Lambda$CDM linear power spectrum.

In order to perform such a large simulation on the available hardware,
a special version of the {\small GADGET-2} code \citep{Springel2001b,
Springel2005Gadget2} was created with very low memory consumption. The
computational algorithm combines a hierarchical multipole expansion,
or `tree' method \citep{Barnes1986}, with a Fourier transform
particle-mesh method \citep{Hockney1981}. The short-range
gravitational force law is softened on comoving scale $5\,h^{-1}{\rm
kpc}$ which may be taken as the spatial resolution limit of the
calculation, thus achieving a dynamic range of $10^5$ in 3D. Data from
the simulation were stored at 63 epochs spaced approximately
logarithmically in time at early times and approximately linearly in
time at late times (with $\Delta t \sim 300$Myr). Post-processing
software identified all resolved dark haloes and their subhaloes in
each of these outputs and then linked them together between
neighboring outputs to construct a detailed formation tree for every
object present at the final time. Galaxy formation modelling is then
carried out in post-processing on this stored data structure.

\subsection{The basic semi-analytic model}
\label{sam}
Our semi-analytic model is that of \citet{Croton2006} as updated by
\citet{Delucia2006b} and made public on the Millenium Simulation data
download site \citep[see][]{Lemson2006}. These models include the
physical processes and modelling techniques originally introduced by
\citet{White1991,Kauffmann1993,Kauffmann1998,Kauffmann1999,
  Kauffmann2000,Springel2001} and \citet{Delucia2004}, principally gas
cooling, star formation, chemical and hydrodynamic feedback from
supernovae, stellar population synthesis modelling of photometric
evolution and growth of supermassive black holes by accretion and
merging. They also include a treatment \citep[based on that
of][]{Kravtsov2004} of the suppression of infall onto dwarf galaxies
as consequence of reionisation heating. More importantly, they include
an entirely new treatment of ``radio mode'' feedback from galaxies at
the centres of groups and clusters containing a static hot gas
atmosphere. The equations specifying the various aspects of the model
and the specific parameter choices made are listed in
\citet{Croton2006} and \citet{Delucia2006b}. The only change made here
is in the dust model as described in the next section.

\subsection{Improved dust treatment for the fiducial model}
\label{subsec:modelext}

Even at low redshifts, a crucial ingredient in estimating appropriate
magnitudes for model galaxies, particularly in the $B$-band, is the
dust model.  For the present-day luminosity function (LF) a simple
phenomenological treatment calibrated using observations in the local
universe has traditionally given satisfactory results
\citep{Kauffmann1999}. However, the situation at high redshift is more
delicate because of the much higher predicted gas (and thus dust)
columns, the highly variable predicted metallicities, and the shorter
emitted wavelengths corresponding to typical observed photometric
bands. We found we had to adopt a new approach in order to be
consistent with current data on extinction in high redshift
galaxies. \citet{Devriendt1999} advocate a dust model based on the HI
column density in the galaxy disk, a quantity that can be estimated
from the cold gas mass and the disk size of a galaxy, both of which
are available for each galaxy in our semi-analytic model. A plausible
scaling of dust-to-gas ratio with metallicity can easily be
incorporated using the metal content given by a chemical evolution
model \citep[cf.][]{Devriendt2000}. Based on this we get
\begin{equation}
  \label{eq:taulambda}
  \tau_\lambda^Z=\left(\frac{A}{A_V}\right)_{Z_\odot}\eta_{Z}
\left(\frac{\langle N_H\rangle}{2.1\times 10^{21} cm^{-2}}\right)  
\end{equation}
with the average hydrogen column density obtained from
\begin{equation}
  \label{eq:nh_mean}
\langle N_H\rangle=\frac{M_{\rm gas}}{1.4\mu m_p \pi r_t^2}~.
\end{equation}
${A}/{A_V}$ here is the extinction curve from \citet{Cardelli1989}.
We assume the dust-to-gas ratio to scale with metallicity and redshift
as $\eta_{Z}=(1+z)^{-\frac{1}{2}}\left({Z_{\rm
gas}}/{Z_\odot}\right)^s$, where $s=1.35$ for $\lambda<2000\,{\rm\AA}$
and $s=1.6$ for $\lambda>2000\,{\rm\AA}$. The factor of
$(1+z)^{-\frac{1}{2}}$ in this formula is adopted in order to
reproduce results for Lyman-break galaxies at $z\sim3$.
\citet{Adelberger2000} find $\langle\tau\rangle_{1600}\lesssim 2$ at
rest-frame $1600\,{\rm\AA}$, showing that dust-to-gas ratios are lower
at this redshift compared to the local universe for objects of the
same $L_{\rm bol}$ and metallicity \citep[a result echoed
in][]{Reddy2006}. This behaviour also agrees with a recent study of
the dust-to-gas/dust-to-metallicity ratio by \citet{Inoue2003}.
Please note that the average extinction of our model galaxies still
increases strongly with redshift due both to the ever shorter
rest-frame bands we probe and to the smaller disk sizes we predict at
higher redshift (see equations \ref{eq:taulambda} and \ref{eq:nh_mean}
above).

\subsection{Making mock observations: lightcones}
\label{subsec:lightcones}

From a theoretical point of view it would be most convenient to
compare predictions for the basic physical properties of galaxies
directly with observation, but in practice this is rarely
possible. For faint and distant object the most observationally
accessible properties are usually fluxes in specific observer-defined
bands.  Quantities such as stellar mass or star-formation rate (often
even redshift) must be derived from these quantities and are subject
to substantial uncertainties stemming primarily from the assumptions
on which the conversion is based.  Moreover which galaxies can be
observed at all (and so are included in observational samples) is
typically controlled by observational selection effects on
apparent magnitude, colour, surface brightness, proximity to other
images and so on.

In order to minimise these uncertainties when drawing astrophysical
conclusions about the galaxy population, it is beneficial to have a
simulated set of galaxies with known intrinsic properties from which
``observational'' properties can be calculated, and to apply the same
conversions and selection effects to this mock sample as to the real
data.  One can then assess the accuracy with which the underlying
physical properties can be inferred. In this approach the uncertain
relations between fundamental and observable quantities become part of
the model, and their influence on any conclusions drawn can be
assessed by varying the corresponding assumptions throughout their
physically plausible range.  A disadvantage is that shortcomings in,
for example, the galaxy formation model are convolved with many other
effects (for example the conversion from mass to luminosity) and
separation of these effects can be difficult. In particular, it may
become difficult to identify why a particular model disagrees with the
data, since effects from many different sources may be degenerate.

We make mock observations of our artificial universe, constructed from
the Millennium Simulation, by positioning a virtual observer at zero
redshift and finding those galaxies which lie on his backward light
cone. The backward light cone is defined as the set of all light-like
worldlines intersecting the position of the observer at redshift
zero. It is thus a three-dimensional hypersurface in four-dimensional
space-time satisfying the condition that light emitted from every
point is received by the observer now. Its space-like projection is
the volume within the observer's current particle horizon.  From this
sphere, which would correspond to an all-sky observation, we cut out a
wedge defined by the assumed field-of-view of our mock observation. It
is common practice to use the term {\it light cone} for this wedge
rather than for the full (all-sky) light cone, and we will follow this
terminology here.

The issues which arise in constructing such light cones have been
addressed in considerable detail by \citet{Blaizot2005}.  In the
following we adopt their proposed solutions in some cases (for
example, when interpolating the photometric properties of galaxies to
redshifts for which the data were not stored) and alternative
solutions in others (for example when dealing with the limitations
arising from the finite extent of the simulation). We refer readers to
their paper for further discussion and for illustration of the size of
the artifacts which can result from the limitations of this
construction process.

There are two major problems to address when constructing a light cone
from the numerical data. The first arises because the Millennium
Simulation was carried out in a cubic region of side $500\mpch$
whereas the comoving distance along the past light cone to redshift 1 
is $2390\mpch$ and to redshift 6 is $6130\mpch$.
Thus deep light cones must use the underlying periodicity and
traverse the fundamental simulation volume a number of times. Care is
needed to minimise multiple appearances of individual objects, and to
ensure that when they do occur they are at widely different redshifts
and are at different positions on the virtual sky. The second problem
arises because redshift varies continuously along the past light
cone whereas we have stored the positions, velocities and properties
of our galaxies (and of the associated dark matter) only at a finite
set of redshifts spaced at approximately 300 Myr intervals out to
$z=1$ and progressively closer at higher redshift. We now present
our adopted solutions to each of these problems in turn.

\subsubsection{How to avoid making a kaleidoscope}
\label{subsubsec:lightconetec}

The underlying scale of the Millennium Simulation $500\mpch$,
corresponds to the comoving distance to $z\sim 0.17$.  However, we
want to produce galaxy catalogues which are at least as deep as the
current observations, and, in practice, to be one or two generations in
advance. Although the periodicity of the simulation allows us to fill
space with any required number of replications of the fundamental
volume, this leads to obvious artifacts if the simulation is viewed
along one of its preferred axes.  We can avoid this kaleidoscopic
effect by orienting the survey field appropriately on the
virtual sky with respect to the three directions defined by the sides
of the fundamental cube. The ``best'' choice depends both on the shape
and depth of the survey being simulated, and on the criteria adopted
to judge the seriousness of the artifacts to be minimised. Here we do
not give an optimal solution to the general problem, but rather a
solution which works acceptably well for deep surveys of relatively
small fields. 

Consider a cartesian coordinate system with origin at one corner of
the fundamental cube and with axes parallel to its sides. Consider the
line-of-sight from this origin passing through the point $(L/m, L/n,
L)$ where $m$ and $n$ are integers with no common factor and L is the
side of the cube. This line-of-sight will first pass through a
periodic image of the origin at the point $(nL, mL, nmL)$, i.e. after
passing through $nm$ replications of the simulation. If we take the
observational field to be defined by the lines-of sight to the four
points $((n\pm0.5/m)L, (m\pm0.5/n)L, nmL)$, it will be almost
rectangular and it will have total volume $L^3/3$ out to distance
$(n^2 + m^2 + n^2m^2)^{0.5}L$.  Furthermore no point of the
fundamental cube is imaged more than once.  This geometry thus gives a
mock light cone for a near-rectangular survey of size $1/m^2n \times
1/n^2m$ (in radians) with the first duplicate point at distance $\sim mnL$.
For example, if we take $m=2$ and $n=3$ we can make a mock light cone
for a $4.8^{\rm o} \times 3.2^{\rm o}$ field out to $z=1.37$ without
any duplications. For $m=3$ and $n=4$ we can do the same for a
$1.6^{\rm o} \times 1.2^{\rm o}$ area out to $z=5.6$. Choosing $m=1$
and $n=5$ results in a $11.5^{\rm o} \times 2.3^{\rm o}$ survey with
no duplications out to $z=1.06$.
 
If we wish to construct a mock survey for a larger field or to a
greater distance than these numbers allow, then we have to live with
some replication of structure.  Choosing the central line-of-sight of
to be in a ``slanted'' direction of the kind just described with $m$
and $n$ values matched roughly to the shape of the desired field
usually results in large separations of duplicates in angle and/or in
redshift. Careful optimisation is needed for any specific survey
geometry in order to get the best possible results. Note that any
point within the fundamental cube can be chosen as the origin of a
mock survey, and that, in addition, there are four equivalent central
lines-of sight around each of the three principal directions of the
simulation.  It is thus possible to make quite a number of equivalent
mock surveys of a given geometry and so to ensure that the full
statistical power of the Millennium Run is harnessed when estimating
statistics from these mock surveys.

Taking into account the above considerations, we select the central
line-of sight to be in the direction of the unit vector ${\bf u}_3$
defined by
\begin{equation}
(m^2 + n^2 + m^2n^2)^{1/2} {\bf u}_3 = (n, m, mn),
\end{equation}
we define a second unit vector ${\bf u}_1$ to be perpendicular
both to ${\bf u}_3$ and to the unit vector along the coordinate 
direction associated with the smaller of $m$ and $n$ (the $x$-axis
in the above examples) and we take a third unit vector ${\bf u}_2$ 
to be perpendicular to the first two so as to define a right-handed
cartesian system. If we define $\alpha$ and $\delta$ as local angular
coordinates on the sky in the directions of ${\bf u}_1$ and ${\bf
u}_2$ respectively, with origin in  our chosen central direction, 
then a particular 3-dimensional position ${\bf x}$ corresponds to 
\begin{eqnarray*}
\tan \alpha =   {\bf x}\cdot{\bf u}_1 / {\bf x}\cdot{\bf u}_3\\
\tan \delta =   {\bf x}\cdot{\bf u}_2 / {\bf x}\cdot{\bf u}_3
\end{eqnarray*}
The position {\bf x} lies within our target rectangular field provided
\begin{eqnarray*}
 \label{eq:conecond}
 |\tan \alpha| \leq  \tan \Delta\alpha/2\\
 |\tan \delta| \leq  \tan \Delta\delta/2\, ,
\end{eqnarray*}
Where $\Delta\alpha$ and $\Delta\delta$ give desired angular extent of
the field in the two orthogonal directions (with
$\Delta\alpha\geq\Delta\delta$ assumed here).  Note that this
formulation of the condition to be within the light cone does not
require any transcendental functions to be applied to the galaxy
positions, allowing membership to be evaluated efficiently. This can
be a significant computional advantage when one is required to loop
over many replications of the (already large) Millennium galaxy
catalogues.

We point out in passing that only for comoving coordinates within a
flat universe do we have the luxury of cutting out light cones from
our (replicated) simulation volume simply as we would in Euclidian
geometry. In general this is a much less trivial endeavour that
requires accounting for the curvature of the universe as well as its
expansion with time. (In addition, second-order effects like
gravitational lensing should, in principle, be taken into account for
any geometry.)

\subsubsection{How to get seamless transitions between snapshots}
\label{subsubsec:lightconetec2}

After determining the observer position and survey geometry we fill
three-dimensional Euclidian space-time with a periodically replicated
grid of simulation boxes, keeping only those which intersect our
survey. In practice, since the Millennium Simulation data at each time
are stored in a set of 512 spatially disjoint cells, we keep only
those cells which intersect the survey. In principle, a galaxy within
our survey at comoving distance $D$ from the observer should be seen
as it was at redshift $z$ where
\begin{equation}
 \label{eq:comovdist}
  D(z)=\int_0^z\frac{cdz'}{H_0\sqrt{\Omega_M(1+z')^3+\Omega_\Lambda}}.
\end{equation}
A problem arises, however, because the positions, velocities and
physical properties of our galaxies are stored only at a discrete
set of redshifts $z_i$ corresponding to a discrete set of distances
$D_i$. (For definiteness we adopt $z_1 = D_1 = 0$ and $z_i>z_{i-1},
D_i > D_{i-1}$.) The comoving distance between outputs is 80
to 240\mpch, corresponding to 100 to 380 Myr, depending on redshift.

One way to deal with this problem would be to interpolate the
positions, velocities and physical properties of the galaxies at each
distance $D$ from the output redshifts which bracket it, e.g. $z_i$
and $z_{i+1}$ where $D_{i+1} > D > D_i$.  We decided against this
procedure for several reasons. In the first place, the Millennium
Simulation appears to give dynamically consistent results for the
galaxy distribution down to scales of 10kpc or so \citep[see, for example,
the 2-point correlation functions in][]{Springel2006}. On such scales characteristic orbital timescales are smaller
than the spacing between our outputs, so interpolation would produce
dynamically incorrect velocities and would diffuse structures. In
addition, the physical properties of the galaxies are not easily
interpolated because of impulsive processes such as mergers and
starbursts. Rather than interpolating, we have chosen to assign the
positions, velocities and physical properties stored at redshift $z_i$
to all survey galaxies with distances from the observer in the range
$(D_i + D_{i+1})/2>D>(D_i + D_{i-1})/2$. Individual small scale
structures are then dynamically consistent throughout this range, and
the physical properties of the galaxies are offset in time from the
correct values by at most half of the time spacing between outputs.

After coarsely filling the volume around the observed light cone with
simulation cells in this way one can simply chisel off the protruding
material, i.e. drop all galaxies which do not lie in the field
according to the condition in \eq\ref{eq:conecond} or which don't
satisfy $(D_i + D_{i+1})/2>D>(D_i + D_{i-1})/2$. The latter condition
causes an additional difficulty since galaxies move between snapshots
and thus it can happen that a galaxy traverses the imaginary boundary
$(D_i + D_{i+1})/2$ between the times corresponding to $z_{i+1}$ and
$z_i$. This results in this galaxy being observed either twice or not
at all, depending on the direction of its motion. We overcome this
problem for galaxies close to the boundary by linearly interpolating
their positions between $z_{i+1}$ and $z_i$ in order to get estimated
positions at the redshift corresponding to $(D_i + D_{i+1})/2$. Those
galaxies whose estimated positions are on the low redshift side of the
boundary are assigned properties corresponding to $z_i$, those on the
high redshift side properties corresponding to $z_{i+1}$.
 
\subsubsection{Getting the right magnitudes}
\label{subsubsec:lightconereq}
The observed properties of a galaxy depend not only on its intrinsic
physical properties but also on the redshift at which it is
observed. In particular, the apparent magnitudes of galaxies are
usually measured through a filter with fixed transmission curve in the
observer's frame. This transmission curve must be blue-shifted 
to each galaxy's redshift and then convolved with the galaxy's
spectral energy distribution in order to obtain an absolute luminosity
which can be divided by the square of the luminosity distance to
obtain the observed flux. A difficulty arises because quantities like
absolute luminosities are accumulated, based on the prior star
formation history of each object, at the time the semi-analytic
simulation is carried out, and they are stored in files which give the
properties of every galaxy at each output redshift $z_i$.  At this
stage the light cone surveys are not yet defined, so we do not know
the redshift at which any particular galaxy will be observed in a
particular mock survey.  We are thus unable to define the filter
function through which its luminosity should be accumulated in order
to reproduce properly the desired observer-frame band.

We deal with this problem in the way suggested by \citet{Blaizot2005}.
We define ahead of time the observer frame magnitudes we wish to
predict, for example, Johnson $B$.  When carrying out the
semi-analytic simulation we then accumulate for all galaxies at
redshift $z_i$ not only the absolute magnitude through the $B$-filter
blue-shifted to $z_i$ but also those for the same filter shifted to
the frequency bands corresponding to $z_{i-1}$ and $z_{i+1}$.  For
galaxies in our mock survey whose physical properties correspond to
$z_i$ but which appear on the light cone at $z>z_i$ we linearly
interpolate an estimate for the observer-frame $B$ absolute magnitude
(at redshift $z$) between the values stored for filters blue-shifted
to $z_i$ and $z_{i+1}$. Similarly, for those similar galaxies which
appear at $z<z_i$ we interpolate the absolute magnitude between the
values stored for filters blue-shifted to $z_i$ and $z_{i-1}$.  It
turns out that this interpolation is quite important. Without it,
discontinuities in density are readily apparent in the distribution of
simulated galaxies in the observed colour-apparent magnitude plane.

\begin{figure*}
\begin{center}
\includegraphics[width=\linewidth]{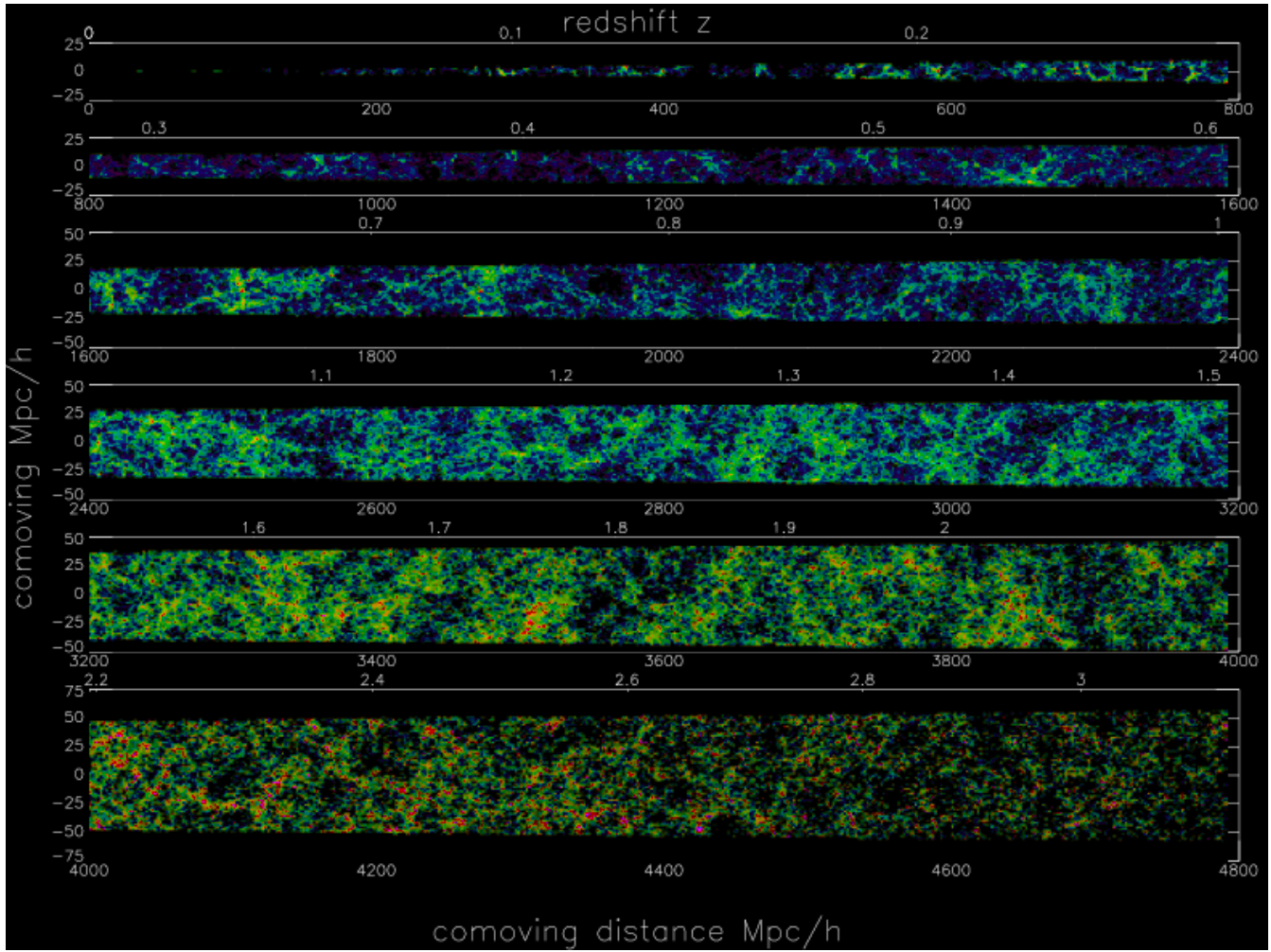}
\caption{\label{lightconefig}Light cone for a $1.4^\circ\times
  1.4^\circ$ field out to $z=3.2$. All galaxies above an apparent
  magnitude limit $K_s(AB)<24$ are shown, where intensity corresponds
  to the logarithmic density and the color denotes the offset
  from the evolving red sequence.}
\end{center}
\end{figure*}

We conclude this chapter by presenting in \fig\ref{lightconefig} an
illustrative example, the simulated light cone of a deep survey (to
$K_s(AB)<24$) of a $1.4^{\rm o}\times 1.4^{\rm o}$ field out to
$z=3.2$. Here intensity corresponds to the logarithmic density and the
colour encodes the offset from the evolving red sequence at the
redshift of observation (assuming passive evolution after a single
burst at $z=6$). Large-scale structure is evident and is well sampled
out to redshifts of at least $z\simeq3$ and it is interesting that at
$z>2$ the reddest galaxies are predicted to be in the densest regions
even though, as we see below, many of them are predicted to be dusty
strongly star forming objects. Individual bright galaxies are
predicted to be visible out to $z\simeq5$ in the full light cone.

\section{Results}
\label{sec:results}

In this section we first compare our model to directly measured
properties of real samples such as their distribution in apparent
magnitude and redshift. We then consider derived properties which
require an increasing number of additional assumptions, moving from
the evolution of rest-frame luminosity functions to that of stellar
mass distributions. Finally we illustrate the large changes predicted
for the distribution of galaxies in rest-frame colour and absolute
magnitude over the redshift range $0<z<3$. This gives a good
impression of the interplay between the various mechanisms that
determine the luminosity and colour of galaxies in our model.

All magnitudes are in the $AB$ system (rather than Vega) unless stated
otherwise.

\subsection{Number Counts}
\label{subsec:counts}
\begin{figure}
\begin{center}
\includegraphics[width=\linewidth]{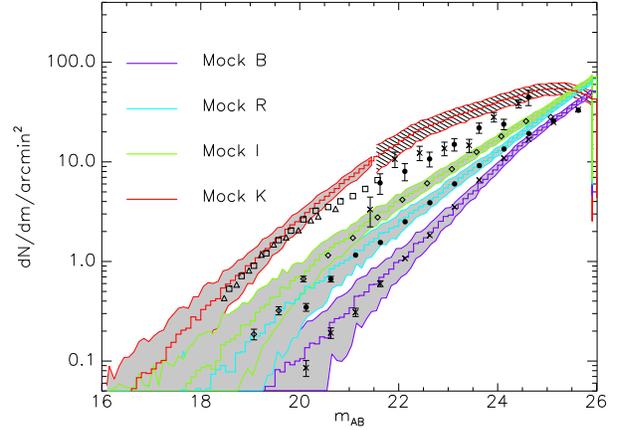}
\caption{\label{brikcountsfig}Predicted galaxy counts per unit area in
four bands compared to a survey in the HDF-N direction
($0.2\,\sq^\circ$) in $BRI$ and to a number of ``wide field'' surveys
at $K_s$ (total of $320$ arcmin$^2$), as well as to deep observations
at $K_s$ in the CDF and HDF-S directions (6 and 7.5 arcmin$^2$
respectively). The grey shaded error bands show $1\sigma$
field-to-field variations assuming an area of $100$ arcmin$^2$,
whereas the hatched error bands show the expected variations for a
smaller square field of area $\sim11$ arcmin$^2$.}
\end{center}
\end{figure}

In \fig\ref{brikcountsfig} we compare predicted galaxy counts obtained
from a mock survey of a $2\,\sq^\circ$ area to observational counts
from a number of different surveys. In the $BRI$ bands we use counts
over a $0.2\,\sq^\circ$ area in the HDF-N direction by
\citet{Capak2004}. In the $K_s$ band we use both the ``wide'' area
($320$ arcmin$^2$ distributed over various fields) counts of
\citet{Kong2006} and the deeper, but smaller area counts in the CDF
and HDF-S directions (6 and 7.5 arcmin$^2$ respectively) by
\citet{Saracco2001}. It is worth noting that for the $BRI$ bands we
were able to use the filter transmission curves appropriate for the
\emph{Subaru} survey, whereas for the $K$ band different effective
transmission curves apply for the different surveys and we have not
taken this into account. In order to quantify the effect of ``cosmic
variance'' (the fact that large statistical fluctuations are expected
in surveys of this size not only from counting statistics but also
from large-scale structure along the line-of-sight) we split up our
$2\,\sq^\circ$ mock survey into 72 fields of size $100$ arcmin$^2$.
The $1\sigma$ scatter among counts in these different areas is shown
as a grey shaded area surrounding the predicted means for $BRI$ and
for the brighter $K$ magnitudes. For the fainter $K$ magnitudes we
split our mock survey into smaller subfields, each with an area of
$\sim11$ arcmin$^2$. The $1\sigma$ variations among these subfields
are shown by the hatched band surrounding the predicted $K$ counts at
fainter magnitudes. Note that this procedure may still somewhat
underestimate the cosmic variance since the different subfields are
not truly independent, but all lie within a single $1.4^{\rm o}\times
1.4^{\rm o}$ mock survey.

In the light of this limitation, and keeping in mind that our
dust-model is still rather simple, it is quite surprising to see the
excellent agreement of the data with our predictions in all three
optical bands.  Agreement at $K$ is less good, and there appears to be
a significant disrepancy faintward of $K_{AB} \sim 21$.  The model
predicts almost twice as many galaxies as are observed at $K_{AB} \sim
23$, although the agreement is again acceptable at $K_{AB} \sim
24.5$. This disagreement appears well outside the statistical errors,
but it should be borne in mind that $K$ magnitudes are extremely
difficult to measure at such faint levels, and it is possible that the
measured quantity does not correspond to the total magnitude assumed
in our modelling.

\subsection{Redshift Distributions for \mbi{K}-selected samples}
\label{subsec:kbandzdist}

\begin{figure}
\begin{center}
\includegraphics[width=\linewidth]{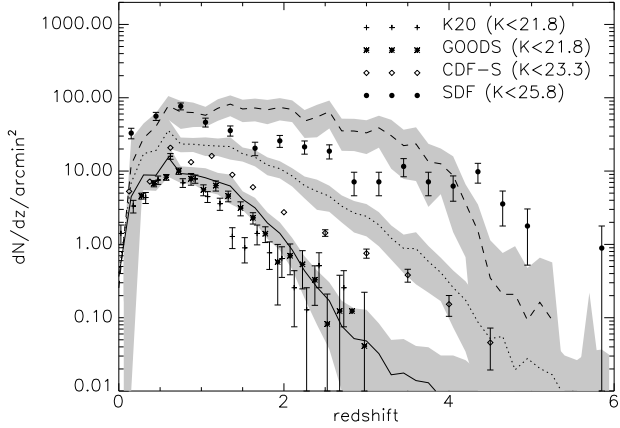}
\caption{\label{k20zdistfig}Predicted redshift distributions for
galaxies to a magnitude limit of $K\le21.8$ (solid), 23.3 (dotted),
and 25.8 (dashed). These are compared to observational results from
K20 and GOODS ($K\le21.8$), from CDF-S ($K\le23.3$) and from SDF
($K\le25.8$).  The latter two are derived purely from photometric
redshift estimates. Error bars on the observational points are based
on counting statistics only.  Grey shaded areas indicate the $1\sigma$
field-to-field scatter assuming an area of $100$ arcmin$^2$ for the
two brighter magnitude limits and $4$ arcmin$^2$ for $K\le25.8$.}
\end{center}
\end{figure}

In \fig\ref{k20zdistfig} we give the redshift distributions predicted
for apparent magnitude limited galaxy samples complete for $K\le21.8$,
$K\le23.3$, and $K\le25.8$. We compare the first of these to data for
a 52 arcmin$^2$ field from K20 \citep{Cimatti2002c} and for a
160 arcmin$^2$ overlapping field from GOODS
\citep{Mobasher2004}. (Note that the name K20 comes from the
survey limit in the Vega system. The two systems are approximately
related by $K_{AB}=K_{\rm Vega}+1.83$.). At the intermediate depth we
compare to the photometric redshift distribution obtained by
\citet{Caputi2006} for a 131 arcmin$^2$ field in the direction of the
Chandra Deep Field South (CDF-S). For the faintest magnitude limit we
compare to the photo-$z$ distribution of a much smaller 4 arcmin$^2$
area in the Subaru Deep Field, as obtained by \citet{Kashikawa2003}.
Again we split up our simulated field into sub-fields of size $100$
arcmin$^2$ ($4$ arcmin$^2$ for $K\le24$) in order to get an estimate
of the expected $1\sigma$ scatter, which we indicate by grey shaded
areas.  For these small fields cosmic variance is quite substantial
and the counts can be influenced significantly by individual galaxy
clusters. This effect is clearly visible in the K20 and GOODS data,
where a pronounced spike is present at $z\simeq0.7$. In addition,
systematic problems with the photometric redshift determinations might
distort the redshift distributions in some ranges.

Despite these uncertainties, our model predictions appear somewhat
high over the redshift range $0.5< z< 1.5$ for the K20 and GOODS
samples. The deeper $K\le23.3$ observations are overpredicted by a
factor of 2 to 3 over the range $1< z <3$. For the faintest sample
there is an apparent overprediction by a somewhat smaller factor over
this same redshift range. Comparing with \fig\ref{brikcountsfig}, we
see that the total overprediction at each of these magnitudes is
consistent with that seen in the counts themselves, although it should
be borne in mind that the CDF-S field is common to both datasets. The
differences we find are larger than the predicted cosmic variance so
they presumably indicate problems with the model (incorrect physics?)
with the observational data (systematics in the magnitudes or
photo-$z$'s) or both.

\subsection{Luminosity Function evolution}
\label{subsec:lfevol}
\citet{Croton2006} demonstrated that at $z=0$ the luminosity
function (LF) for our model agrees well with observation both in $b_J$
and in $K$. Splitting galaxies according to their intrinsic colours,
these authors also found quite good fits to the LF's for red and blue
galaxies separately, with some discrepancies for faint red
galaxies. Here we compare the evolution of the LF predicted by our
model in {\it rest-frame} $B$ and $K$ band with recent observational
results.

\subsubsection{The B-band Luminosity Function}
\label{subsubsec:bbandlf}

\begin{figure}
\begin{center}
\includegraphics[width=\linewidth]{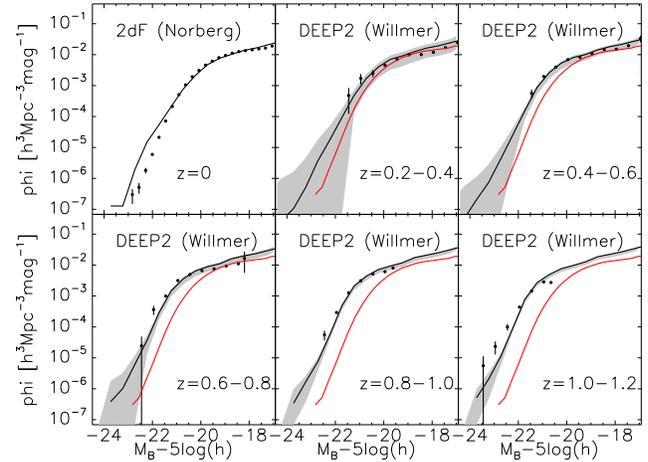}
\caption{\label{blf1fig}Comparison of the rest-frame $B$-band LF
predicted by our simulation for redshifts in the range $0<z<1.2$ to
observational estimates from \citet{Norberg2002} at $z=0$ and from
\citet{Willmer2005} at higher redshifts. The local LF of the upper
left panel is repeated as a thin red line in each of the other
panels. A grey shaded region surrounding each model prediction shows
the $1\sigma$ scatter expected in observational estimates based on
samples similar in size to the corresponding observational sample.}
\end{center}
\end{figure}

In \fig\ref{blf1fig} we compare the evolution of the rest frame $B$-band
LF predicted by our simulation to results from the DEEP2 survey
\citep{Willmer2005}. As a $z=0$ standard we use the local LF from the
2dF survey \citet{Norberg2002}. This is compared with our model in the
top-left panel and is repeated as a thin red line in each of the other
panels, where the high redshift data are indicated by points with
error bars. Our predicted LF is shown in each panel as a solid line
with a grey area indicating the $1\sigma$ scatter to be expected for
an estimate from a survey similar in effective volume to the
observational survey. (Note that in all cases the Millennium
Simulation is much larger than this effective volume, so that counting
noise uncertainties in the prediction are negligible.) 

At $z=0$ the agreement between model and observation is excellent.
This is a consequence of the fact that \citet{Croton2006} and
\citet{Delucia2006a} adjusted model parameters in order to optimise
this agreement. However, over the full redshift range from $z=0.2$ to
$1.2$ the predicted LF's agree with the DEEP2 data at the $1\sigma$
level or better. On closer examination, it appears that the model
somewhat overpredicts the observational abundance fainter than the
knee of the luminosity function, by a factor $\sim 1.5$ depending on
redshift.  On the other hand, at the higher redshifts very luminous
galaxies appear slightly more abundant in the real data than in the
model.  It is important to keep in mind that our dust model has a
strong influence here, and plausible modifications to it might account
for either or both of these minor discrepancies. In general, the
agreement with the data seems quite impressive, at least in this band
and over this redshift range.

\subsubsection{The K-band Luminosity Function}
\label{subsubsec:kbandlf}

Model predictions for the rest-frame $K$-band LF should, in
principle, be more robust than predictions for the rest-frame $B$-band,
because the effects of our uncertain dust modelling are then much
weaker. On the other hand, observational determinations of the LF at
rest-frame $K$ are more uncertain than at rest-frame $B$, because
the magnitudes of high redshift galaxies must then be inferred by
extrapolation beyond the wavelength region directly measured, rather
than interpolated between the observed bands.  This situation is
improving rapidly as deep data at wavelengths beyond 2$\mu$ become
available from {\it Spitzer}.

\begin{figure}
\begin{center}
\includegraphics[width=\linewidth]{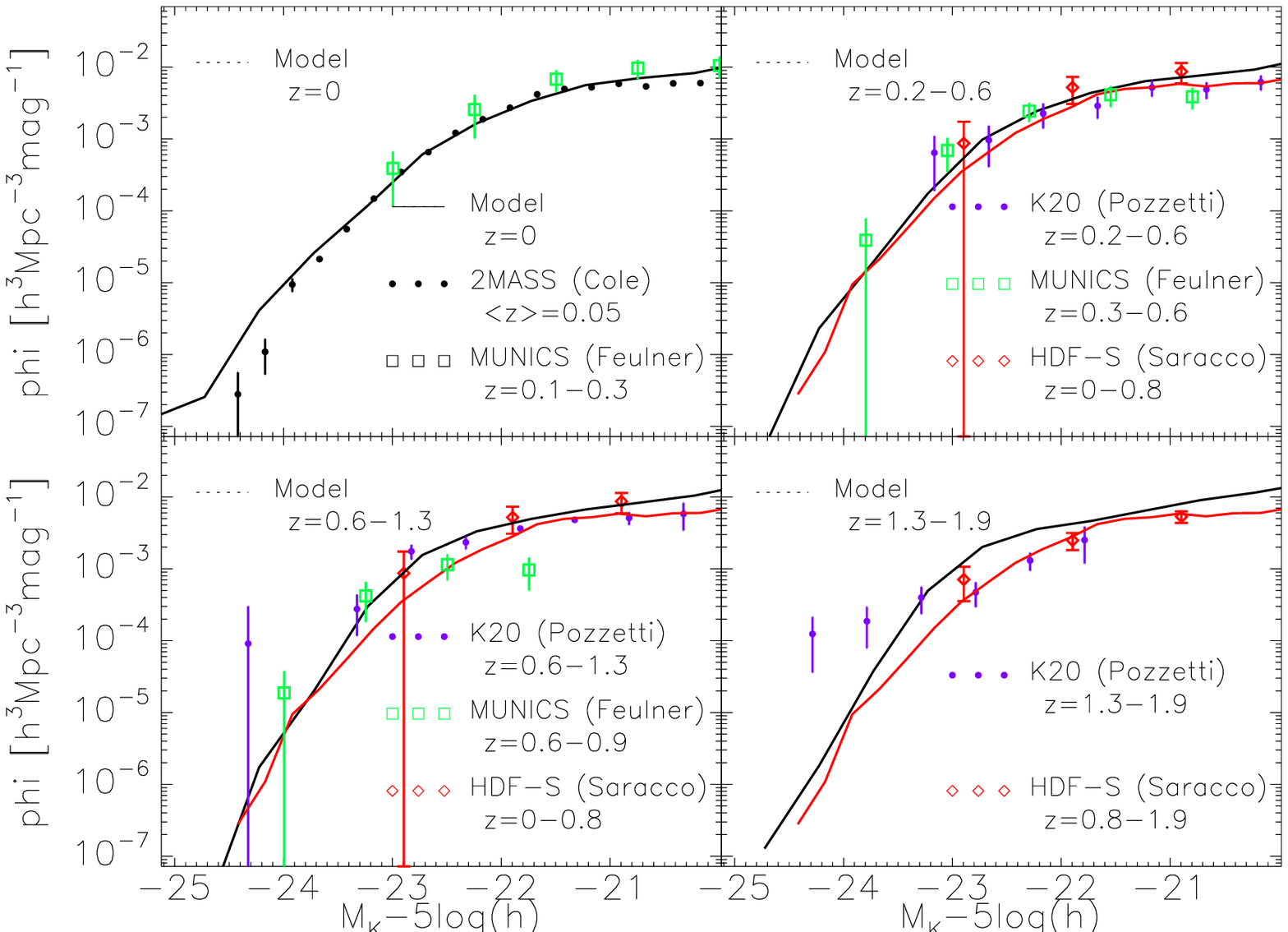}
\caption{\label{klf1fig}Comparison of the evolution predicted for the
rest-frame $K$-band LF to observational determinations from
\citet{Cole2001} at low $z$ (upper left panel, repeated as a thin
red line in the other panels) and from \citet{Pozzetti2003},
\citet{Feulner2003} and \citet{Saracco2006} at higher redshifts.}
\end{center}
\end{figure}

As can be seen in \fig\ref{klf1fig}, our predictions for the evolution
of the rest-frame $K$-band LF show the same behaviour as for the
$B$-band. The local result from \citet{Cole2001} is reproduced well,
as illustrated in the upper left panel and already demonstrated in
\citet{Croton2006}. The observed $z=0$ function is reproduced as a
thin red line in the other panels in order to make the amount of
evolution more apparent.  At higher redshifts we compare with
observational determinations from \citet{Pozzetti2003} for the 52
arcmin$^2$ of the K20 survey, from \citet{Feulner2003} for the 600
armin$^2$ of the MUNICS sample and from \citet{Saracco2006} for a 5.5
arcmin$^2$ area in the HDF-S. In these plots we give error bars as
quoted by the original papers, but we note that these are based on
counting statistics only and additional uncertainties are expected due
to clustering, particularly for the smaller fields.  Furthermore,
photo-$z$'s are used for a significant number of galaxies in these
determinations which may lead to additional systematic uncertainties
in the results. Given the scatter between the various observational
determinations, the disagreements between model and data do not look
particularly serious. The models do appear to overpredict the
abundance of galaxies near the knee of the luminosity function,
perhaps by a factor of 2 at the highest redshift, echoing the
discrepancies found above when comparing with $K$-band galaxy counts
and redshift distributions.  

\subsubsection{Evolution of Luminosity Function parameters}
\label{subsubsec:paramlfevol}

\begin{figure}
\begin{center}
\includegraphics[width=\linewidth]{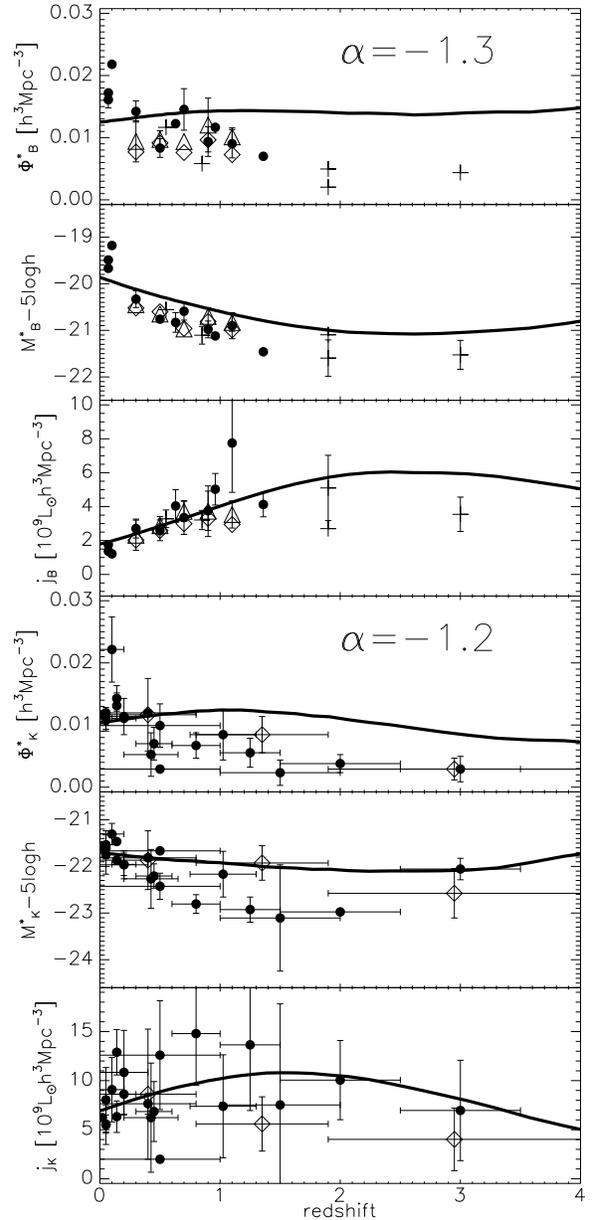}
\caption{\label{blf2fig}The evolution of the $\Phi^*$ and $M^*$
parameters of Schechter (1976) fits to luminosity functions in
rest-frame $B$ and $K$ (with $\alpha$ held constant at the values
indicated). We also show the evolution of the total luminosity density
$j$ inferred from these parameters. In each panel the solid line
denotes the model prediction and the symbols are data from different
sources (with potentially different $\alpha$). The $B$ band data
comprises observations from \citet[+ symbols]{Poli2003} and
\citet[diamonds]{Faber2005}, who also provide a compilation from the
literature (filled circles), whereas the $K$ band data are
observations (diamonds) and a literature compilation (filled circles)
from \citet{Saracco2006}.}
\end{center}
\end{figure}

In order to display the evolution of the luminosity function in our
models more effectively, we have fit Schechter (1976) functions to the
simulation data for the rest-frame $B$ and $K$-bands at every stored
output time. In most cases these functions are a good enough fit to
give a fair representation of the numerical results.  In
\fig\ref{blf2fig} we plot the evolution with redshift of the
parameters $\Phi^*$ and $M^*$ and of the volume luminosity density,
$j=\Phi^*L^*\Gamma(\alpha+2)$ using thick solid lines, and we compare
with fits to observational data. For each observational point in the
$K$-band panels we indicate the (often broad) redshift range to which it
refers by a horizontal bar.  The vertical bar indicates the
uncertainty quoted by the original authors.

Not surprisingly, the results of the last section are confirmed.  For
the model $\Phi^*$ increases slightly with redshift out to about
$z=1.5$, whereas the observations imply a relatively steep decline
over this same redshift range. This holds for both photometric
bands. For $M^*$ we see brightening both in the models and in the
observations, but the effect is more pronounced in the latter. In $K$
the models predict $M^*$ to be almost independent of
redshift. Derivations of $\Phi^*$ and $M^*$ from observational data
using maximum likelihood techniques usually give results where the
errors in the two quantities correlate in a direction almost parallel
to lines of constant luminosity density. For this reason we expect $j$
to be more robustly determined from the data than either $\Phi^*$ or
$M^*$ individually. It is interesting that the apparent deviations
between data and model for $\Phi^*$ and $M^*$ largely compensate, so
that the model predicts an evolution of $j$ which is quite similar to
that inferred from the observations.  This is particularly striking at
rest-frame $B$.  At rest-frame $K$ the observational error bars are
still too large to draw firm conclusions, but a non-evolving
luminosity density represents the data somewhat better than does our
model, again confirming the conclusions we drew in earlier sections.

\subsection{The evolution of the stellar mass function}
\label{subsec:smfevol}

The evolution of the abundance of galaxies as a function of their
stellar mass is one of the most direct predictions of galaxy formation
models.  It depends on the treatment of gas cooling, star-formation
and feedback, but not directly on the luminous properties of the stars
or on the dust modelling. (There remains an indirect dependence on the
latter since observations of galaxy luminosities are typically used to
set uncertain efficiency parameters in the modelling.) The stellar
masses of galaxies can also be inferred relatively robustly from
observational data provided sufficient observational information is
available (e.g. Bell \& de Jong 2001, Kauffmann et al 2003). At high
redshift, however, such inferences become very uncertain unless data
at wavelengths beyond 2$\mu$ are available (e.g. from {\it
Spitzer}). The observationally inferred masses also depend
systematically on the assumed Initial Mass Function for star formation
(usually taken to be universal) and it is important to ensure that
consistent assumptions about the IMF are made when comparing
observation and theory.

\begin{figure}
\begin{center}
\includegraphics[width=\linewidth,height=12cm]{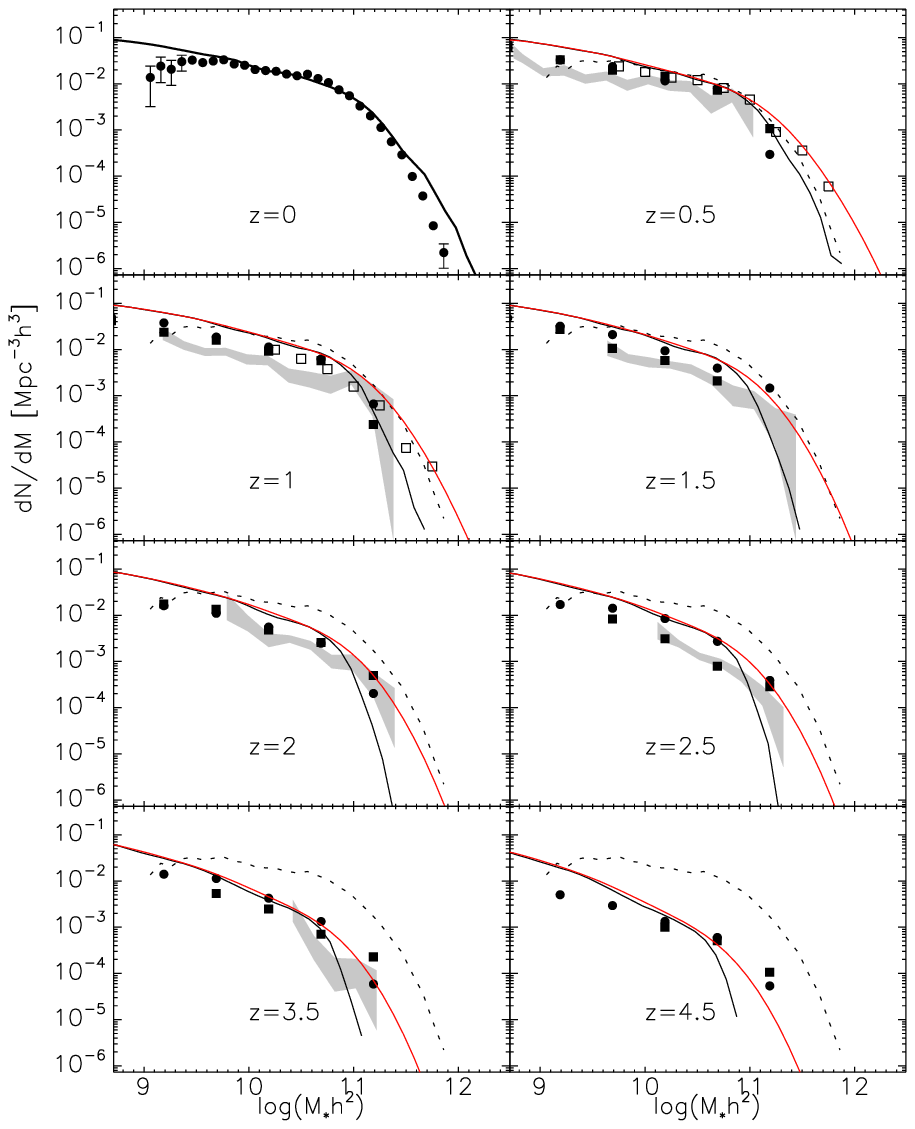}
\caption{\label{smfevolfig}Evolution of the stellar mass function in
  the redshift range $z=0-4.5$. Local data are from \citet{Cole2001}
  and are repeated as a black dashed line in the higher redshift
  panels.  High redshift data are taken from
  \citet[][symbols]{Drory2005} and \citet[][grey shaded
  areas]{Fontana2006}. Model predictions are shown both with (red) and
  without (black) convolution with a normal distribution of standard
  deviation 0.25 representing measurement errors in $\log M_*$.  At
  $z=0$ we consider the mass determinations precise enough to neglect
  this effect.}
\end{center}
\end{figure}

Bearing in mind these caveats, \fig\ref{smfevolfig} compares the mass
functions predicted by our model to local data from \citet{Cole2001}
as well as to high-redshift estimates from \citet{Drory2005}, based on
the MUNICS survey, and from \citet{Fontana2006} based on the
MUSIC-GOODS data.  The latter study uses data in the 3.6 to 8$\mu$
bands from {\it Spitzer} to constrain the spectral energy
distributions of the galaxies and so should give substantially more
reliable results at high redshift than the former.  In
\fig\ref{smfevolfig} the model mass functions at $z>0$ are shown both
before and after convolution with a gaussian in $\log M_*$ with
standard deviation 0.25. This is intended to represent the uncertainty
in the observational determinations of stellar mass. This error may be
appropriate for the MUSIC-GOODS sample at all redshifts, but it is
certainly too small to represent uncertainties in the MUNICS mass
estimates at high redshift. We note that such errors weaken the
apparent strength of the quasi-exponential cut-off at high masses.  We
neglect their effects at $z=0$.

Our model is nicely consistent with the observed mass function in the
local Universe, but it clearly overpredicts the abundance of galaxies
at redshifts between 1 and 3.  The observed evolution relative to the
$z=0$ function (indicated by the dashed line in each of the higher
redshift panels) is strong, while the model prediction is rather more
modest. In the stellar mass range $10^{10}$ to $10^{11}M_\odot$ where
the observational estimats appear most reliable the overprediction
reaches a factor of about 2 at $z=2$. This is nicely consistent with
the conclusions we reached in earlier sections based on number counts,
redshift distributions and luminosity functions, but unfortunately
the scatter between the various observational determinations is large
enough to prevent any firm conclusion.

\subsection{The evolution of the colour-magnitude relation}
\label{subsec:cmrevol}

\begin{figure}
\begin{center}
\includegraphics[width=\linewidth]{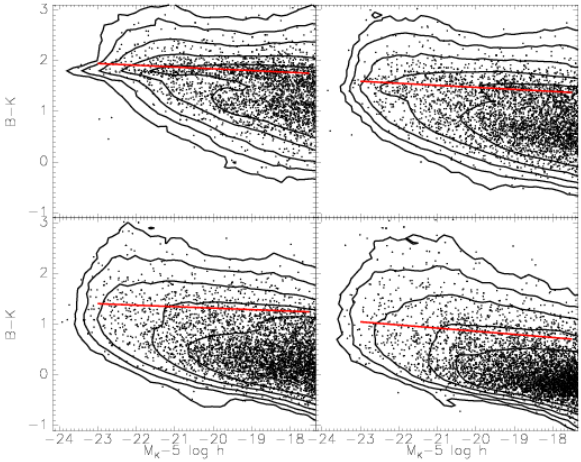}
\caption{\label{kcmevolfig}  Evolution of the colour-magnitude
  distribution of galaxies in rest-frame $B$ and $K$ over the redshift
  range $z=[0,3]$. Randomly selected 10000 galaxies in a
  $2.5\times10^5h^{-3}$Mpc$^3$ volume of the simulation are plotted in
  each panel. For comparison purposes the distribution of a much
  larger sample in a $1.5\times10^7h^{-3}$Mpc$^3$ volume is indicated
  as logarithmically spaced contours. The red solid line indicates the
  colour magnitude relation predicted for stellar populations formed
  in a single burst at $z=6$ and evolving passively thereafter.  The
  metallicity of the populations has been adjusted as a function of
  stellar mass to fit the ridge line of the $z=0$ red sequence.}
\end{center}
\end{figure}

Recent studies of the high redshift galaxy population have often
stressed the presence of massive objects with colours similar to those
expected for fully formed and passively evolving ellipticals
\citep[e.g.][and references therein]{Renzini2006}. This is usually
presented as a potential problem for ``hierarchical'' models of galaxy
formation where star formation and merging continue to play a major
role in the build up of galaxies even at recent times.  In order to
illustrate how these processes are reflected in the colours and
magnitudes of galaxies in our simulation, we show in
\fig\ref{kcmevolfig} the colour-magnitude diagram for 10000 galaxies
randomly sampled from a $2.5\times10^5h^{-3}$Mpc$^3$ volume at
redshifts $z=0,$ 1, 2 and 3. At $z=0$ the well-known bi-modal
distribution of colours is very evident. A tight red-sequence of
passively evolving objects is present with a slope reflecting a
relation between mass and metallicity. There is also a ``blue cloud''
of star-forming systems.  A success of the model emphasised by
\cite{Croton2006} is the fact that the brightest galaxies all lie on
the red sequence at $z=0$.  This is a consequence of including a
treatment of ``radio feedback'' from AGN.

To allow better appreciation of the evolution to high redshift, we
also show logarithmically spaced contours of the colour-magnitude
distribution of all galaxies in a $1.5\times10^7h^{-3}$Mpc$^3$ volume
as black contours in the panels of \fig\ref{kcmevolfig}. The bluing of
the upper envelope with increasing redshift is very clear and is
consistent with passive evolution of the red sequence. We illustrate
this by fitting a population synthesis model to the ridge line of the
$z=0$ red sequence, assuming a single burst of star formation at $z=6$
and a metallicity which varies with stellar mass. This model is shown
as a red line not only at $z=0$ (where it was fit) but also at the
earlier redshifts. Notice that although there are galaxies with red
sequence colours at all redshifts, the sequence becomes less and less
well-defined at earlier times, with a substantial number of objects
appearing {\it redder} than the passively evolving systems.  These are
compact, gas- and metal-rich galaxies where our model predicts very
substantial amounts of reddening. Recent surveys of distant Extremly
Red Objects have found substantial numbers of such systems
\citep{Cimatti2002a,Cimatti2003,LeFevre2005,Kong2006}, but it remains
to be seen if our model can account quantitatively for their
properties.

The colours of galaxies in the blue cloud also become bluer at high
redshift.  This is a consequence of an increase in the typical ratio
of current to past average star formation rate in these galaxies.  The
difference between star-forming systems and ``true'' red-sequence
galaxies becomes blurred at high redshift in our model because of the
increasingly important effects of dust.

\section{Discussion and Conclusions}
\label{sec:discussion}

The model we have used in this paper is that of \citet{Springel2005b}
and \citet{Croton2006} and as updated by \citet{Delucia2006b} and made
public through the Millennium Simulation download site
\citep[see][]{Lemson2006}. Earlier work has compared this model to a
wide range of properties of low redshift galaxies: their luminosity
functions, their bi-modal luminosity-colour-morphology distribution
and their Tully-Fisher relation \citep{Croton2006}; their spatial
clustering as inferred from two-point correlations
\citep{Springel2005b,Li2006,Meyer2006} and from fits to halo
occupation distribution models \citep{Wang2006,Weinmann2006}; their HI
gas content \citep{Meyer2006}; and their assembly histories within
clusters \citep{Delucia2006a,Delucia2006b}. Although
\citet{Croton2006} compared the evolution of the global star formation
rate and the global black hole accretion rate of the model to
observation, the current paper is the first to compare its predictions
in detail with observations of high redshift galaxies.

Our comparison to galaxy counts, to redshift distributions and to
observational estimates of luminosity and mass functions at high
redshift paints a consistent picture despite large statistical
uncertainties and some significant technical issues.  Our model
appears to have too many relatively massive galaxies at high redshift
and these galaxies appear to be too red. Thus, while we fit optical
galaxy counts well up to densities of 30 gal/mag/arcmin$^2$, we start
to overpredict numbers in the $K$ band at densities above about 3
gal/mag/arcmin$^2$. This overabundance of apparently red galaxies
shows up in the redshift distributions as an overprediction of the
number of galaxies with $K\sim23$ to 25 at redshift between about 1
and 3.  These correspond to moderately massive systems near the knee
of the luminosity function, and indeed, while our rest-frame $B$
luminosity functions appear compatible with observation out to $z\sim
1$, at rest-frame $K$ our luminosity functions are noticeably high
beyond $z=0.5$ except possibly for the brightest objects. The problem
shows up most clearly in our mass functions which overpredict
observationally estimated abundances by about a factor of 2 at
$z=2$. Apparently the mass function of galaxies evolved more strongly
in the real Universe than in our simulation.

A galaxy formation model with similar basic ingredients to ours, but
with important differences of detail has been independently
implemented on the Millennium Simulation by \citet{Bower2006}. This
model is also publicly available at the download site. It fits low
redshift galaxy luminosity functions as well as our model, but the
comparisons which \citet{Bower2006} show to high-redshift luminosity
and mass function data (essentially the same datasets we use here)
demonstrate somewhat better agreement than we find in this paper.  In
the mass range $10^{10}< h^2M_*/M_\odot< 10^{11}$ the abundances
predicted by their model are lower than ours by about 20\% at $z=1$
and by about 30\% at both $z=2$ and $z=3.5$, despite the fact that at
$z=0$ the two models agree very well. This is consistent with the fact
that their model forms 20\% of all its stars by $z=3.2$ and 50\% by
$z=1.65$ whereas the corresponding redshifts for our model are $z=3.6$
and $z=1.9$.  These differences arise from details of the star
formation and feedback models adopted in the two cases.

In summary, both the \citet{Bower2006} simulation and our own are
consistent with most current faint galaxy data.  Thus there
seems no difficulty in reconciling the observed properties of distant
objects with hierarchical galaxy formation.  The fact that predictions
from the two simulations differ at a level which can be marginally
separated by the observations, means that currently accessible
properties of distant galaxies can significantly constrain models of
this type, and hence the detailed physics which controls the formation
and the observable properties of galaxies. The fact that the
model  we test here apparently {\it overpredicts} the abundance of
moderately massive galaxies at high redshift, despite the fact that
late merging plays a major role in the build-up of its more massive
galaxies \citep[e.g.][]{Delucia2006a,Delucia2006b}, demonstrates that
current data are still far from constraining the importance of this
process.  As the data improve, the models will have to improve also to
remain consistent with them.  This interplay between theory and
observation should eventually lead to a more convincing and more
complete picture of how galaxies came to take their present forms.

{\bf Acknowledgements:} We thank Jeremy Blaizot, Darren Croton and
Gabriella De Lucia for useful discussions of a number of
technical issues which arose during this project.  MGK acknowledges a
PhD fellowship from the International Max Planck Research School in
Astrophysics, and support from a Marie Curie Host Fellowship for Early
Stage Research Training.

\bibliographystyle{aa}
\bibliography{references}

\end{document}